%% file: Liu_WCL2026-2681.tex
\documentclass[lettersize,journal]{IEEEtran}
\usepackage{amsmath,amsfonts}
\usepackage{algorithmic}
\usepackage{array}
\usepackage[caption=false,font=normalsize,labelfont=sf,textfont=sf]{subfig}
\usepackage{textcomp}
\usepackage{stfloats}
\usepackage{url}
\usepackage{verbatim}
\usepackage{graphicx}
\def\BibTeX{{\rm B\kern-.05em{\sc i\kern-.025em b}\kern-.08em
		T\kern-.1667em\lower.7ex\hbox{E}\kern-.125emX}}
\usepackage{balance}
\usepackage{amssymb}
\usepackage{algorithm}
\usepackage{cite}
\usepackage{bm}
\usepackage{cuted}
\usepackage{placeins}

\usepackage[utf8]{inputenc}
\usepackage{lineno} 
\usepackage{cases}
\usepackage{float}  
\usepackage{textcomp}
\usepackage{bbold}\usepackage{xcolor}
\usepackage{booktabs}
\usepackage{xurl}

\usepackage{siunitx}
\input{sym_marco.tex}

\usepackage{caption}
\usepackage{subcaption}
\usepackage{amsthm}

\begin{document}
	\bstctlcite{BSTcontrol}
	\title{Joint Beamforming and Phase Shifts Design for RIS-Enabled RSMA-ISAC Systems}
	\author{Xuejun Cheng,~\IEEEmembership{Graduate Student Member,~IEEE,}
		Qian Zhang,~\IEEEmembership{Member,~IEEE,}
		Yuhui Jiao,
		Yufei Zhao,~\IEEEmembership{Member,~IEEE,}
		Zheng Dong,~\IEEEmembership{Member,~IEEE,}
		Ju Liu,~\IEEEmembership{Senior Member,~IEEE}
		
		\thanks{
			This research was supported in part by the Shandong Provincial Natural Science Foundation under Grant ZR2023LZH003;
			The corresponding authors: Ju Liu; Zheng Dong. E-mail: \{juliu, zhengdong\}@sdu.edu.cn. }
		\thanks{Xuejun Cheng, Yuhui Jiao, Zheng Dong, and Ju Liu are with School of Information Science and Engineering, Shandong University, Qingdao, 266237, China. (email: \{chengxuejun, yuhuijiao2024\}@mail.sdu.edu.cn; \{zhengdong, juliu\}@sdu.edu.cn.)}
		\thanks{Qian Zhang is with School of Computer and Communication Engineering, Northeastern University at Qinhuangdao, Qinhuangdao 066004, China (e-mail: zhangqian@neuq.edu.cn).}
		\thanks{Yufei Zhao is with School of Electrical and Electronic Engineering, Nanyang Technological University, Singapore 639798 (e-mail: yufei.zhao@ntu.edu.sg).}}
	
	\maketitle
	
	\begin{abstract}

		This paper investigates the sensing-centric design of reconfigurable intelligent surface (RIS)-enabled rate-splitting multiple access-integrated sensing and communication (RSMA-ISAC) systems. Specifically, we propose a new beam-gain approximation method to enhance the sensing beam gain while satisfying communication quality-of-service (QoS) constraints.
		Since the joint optimization of the beamforming vectors and RIS phase shifts is highly coupled and non-convex, existing methods typically rely on generic optimization solvers involving substantial computational complexity. To address this issue, we propose an efficient constraints-separation-based alternating optimization algorithm (CS-AO). 
		Our proposed algorithm effectively decouples the optimization variables and yields closed-form solutions for all subproblems, thereby significantly reducing the computational burden.		
		Simulation results show that the proposed algorithm achieves sensing beam-gain performance comparable to successive convex approximation (SCA) and semidefinite relaxation (SDR) benchmarks, while achieving more than 120-fold and 50-fold runtime reductions.
		In addition, compared with conventional space-division multiple access (SDMA) schemes, the proposed design exhibits substantial sensing beam gain. 
	\end{abstract}
	\vspace{-5pt}
	
	\vspace{-13pt}	
	\section{Introduction}
	\vspace{-5pt}
	\label{s1}
	In recent years, with the rapid advancement of wireless communication technologies, numerous vertical applications have continuously emerged, including urban air mobility, aerial logistics, and intelligent transportation~\cite{11370843}. 
	These applications require both high-throughput data transmission and high-resolution sensing capabilities. 
	Integrated sensing and communication (ISAC) has been regarded as a promising technology to improve spectrum efficiency by sharing wireless hardware and spectrum resources~\cite{9737357}. 
	However, the performance of ISAC systems is essentially constrained by wireless propagation environments, particularly when line-of-sight (LoS) links are obstructed.
	Reconfigurable intelligent surface (RIS) provides an effective solution by reconfiguring wireless channels through dynamic phase-shift adjustment, thereby enhancing communication and sensing performance~\cite{11139112}.

	In ISAC systems, communication and sensing performances are intrinsically coupled, and the resulting mutual interference makes it challenging to optimize both simultaneously.
	To address this issue, existing ISAC frameworks primarily rely on space division multiple access (SDMA) technology to separate communication and sensing signals, which implicitly treats all interference as noise, exhibiting limitations in interference management~\cite{11036692}.
	Fortunately, rate-splitting multiple access (RSMA) can be employed as an effective interference management technique by splitting user messages into common and private streams~\cite{10038476,11455905}. Its unique advantage lies in enabling both interference decoding and treating interference as noise, thereby achieving more flexible interference management.
	
	Building on this advantage, recent research has introduced RSMA into RIS-ISAC systems to enhance performance in coupled communication-and-sensing scenarios.
	In \cite{10812007}, Zhang {\it et al.} jointly designed communication and radar precoders and optimized RIS phase shifts to maximize the minimum secrecy rate in downlink RIS-ISAC systems.
	In \cite{11165755}, Salem {\it et al.} jointly optimized transmit beamforming, artificial noise, RIS phase shifts, and radar receive beamforming to enhance robust security while ensuring sensing performance. 
	Although existing studies have investigated RIS-RSMA-ISAC systems from the perspectives of secure transmission, energy efficiency optimization, and sensing performance enhancement, their solution procedures mostly rely on general-purpose convex optimization tools, such as CVX, resulting in high computational complexity.
	Meanwhile, closed-form or low-complexity beamforming designs have been proposed for conventional ISAC systems without RIS~\cite{9531111}. 
	However, these methods are difficult to directly extend to RIS-aided scenarios, since RIS introduces additional unit-modulus phase-shift constraints and cascaded-channel coupling. 
	Therefore, designing a low-complexity closed-form algorithm for RIS-enabled RSMA-ISAC systems remains challenging.
		
	To tackle these challenges, we propose a sensing-centric design for RIS-enabled RSMA-ISAC systems. Specifically, 
	1) we introduce a new beam-gain approximation method to enhance the sensing beam gain by jointly optimizing the beamforming vectors, RIS phase shifts, and beam phases. 
	2) we develop a constraint-separation-based alternating optimization (CS-AO) algorithm based on the proximal distance algorithm (PDA), to efficiently solve the resulting non-convex optimization problem, and we derive closed-form solutions for all optimization variables.
	3) simulation results demonstrate that the proposed CS-AO algorithm achieves sensing beam-gain comparable to successive convex approximation (SCA) and semidefinite relaxation (SDR) benchmarks with significantly lower runtime, while the proposed design outperforms conventional SDMA schemes in sensing beam gain.
	
	\vspace{-10pt}
	\section{System Model and Problem Formulation}
	\vspace{-3pt}
	\subsection{System Model}
\begin{figure}[t]
		\centering
		\includegraphics[width=0.35\textwidth]{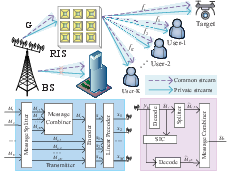}
		\caption{The model of RIS-enabled RSMA-ISAC system.}
		\label{fig_model}
	\end{figure}

	\vspace{-3pt}
	\label{s2}
	We consider a RIS-enabled downlink RSMA-ISAC system, as shown in Fig.~\ref{fig_model}. 
	The system consists of a dual-functional $M$ transmit antennas base station (BS), a RIS equipped with $N=N_1 \times N_2$ reflecting elements, $K$ single-antenna communication users indexed by $\mathcal{K} = \{1, 2, \ldots, K\}$, and a point-like target. 
	
	\textit{1) Communication model:}
	We employ RSMA at the BS to serve multiple communication users. 
	Specifically, the message $M_k$ of the $k$-th user is split into a common part $M_{c,k}$ and a private part $M_{p,k}$. The common parts of all users are jointly encoded into a single common stream $s_c$, while each private part is independently encoded into its corresponding private stream $s_k$. 
	Define $\bm{s} = [s_{1}, \ldots, s_{K}, s_c]^\mathrm{T} \in \mathbb{C}^{(K+1)\times1}$ as the transmitted data stream vector, where $\mathbb{E} \left[\bm s \bm s^\mathrm{H}\right] = \mathbf{I}$. 
	Consequently, the transmitted signal at the BS can be expressed as
	$\bm{x} = \mathbf{W}\bm{s} = \bm{w}_c s_c + \sum_{i\in\mathcal{K}} \bm{w}_i s_{i}$, 
	where $\mathbf{W} = [\bm{w}_1, \ldots, \bm{w}_K, \bm{w}_c] \in \mathbb{C}^{M\times(K+1)}$, $\bm w_c$ and $\bm w_k$ denote the linear precoding vectors for the common stream and the private stream of user-$k$. 
	The direct BS-users links are obstructed.
	Therefore, the received signal $y_k$ at the user-$k$ is expressed as
		\vspace{-7pt}
	\begin{equation} \label{y_K}
		\begin{split}
			y_k =  {\bm{h}_k^\mathrm{H} \bm{w}_c s_c} \ \! + \ \! {\bm{h}_k^\mathrm{H} \bm{w}_k s_k} \ \! + \!\!\!\! {\sum_{i\in\mathcal{K}, i \neq k} \bm{h}_k^\mathrm{H} \bm{w}_i s_i} \!\!\! +  \ \! n_k ,
		\end{split}
	\end{equation}
	\vspace{-13pt}

	\noindent where $\bm h_k^\mathrm{H} = \bm f_k^\mathrm{H} \bm\Theta \mathbf{G}$,  
	$\mathbf{G}\in\mathbb{C}^{N\times M}$ and $\bm f_k\in\mathbb{C}^{N\times 1}$ represent the channels from the BS to RIS, from RIS to user-$k$, respectively.
	$\bm \Theta = \mathrm{diag}\left(\bm \theta^\mathrm{H}\right)\in \mathbb{C}^{N \times N}$ denotes the phase-shift matrix of RIS, where $\bm \theta = [\theta_1, \theta_2, \dots, \theta_N]^\mathrm{T}$, $\theta_n = e^{j\phi_n}$, and $\phi_n \in [0, 2\pi]$.
	$ n_k \sim \mathcal{C} \mathcal{N}\left(0,\sigma_k^2\right)$ denotes the additive white Gaussian noise (AWGN) at the user-$k$. 
	
	Each user initially treats all private streams as interference and then decodes the common stream $s_c$. 
	After successful decoding, $s_c$ is removed by successive interference cancellation (SIC), and user-$k$ then decodes its private stream $s_k$. Hence, the signal-to-interference-plus-noise ratio (SINR) for the common and private streams at user-$k$ are given by
	\vspace{-3pt}
	\begin{equation*}
		\begin{split}
			\gamma_{c,k} = \frac{|\bm h_k^\mathrm{H} \bm w_c|^2}{\sum_{i\in\mathcal{K}}{|\bm h_k^\mathrm{H} \bm w_i|^2}+\sigma_k^2},
			\gamma_{k} = \frac{|\bm h_k^\mathrm{H} \bm w_k|^2}{\sum_{i\in\mathcal{K}, i \neq k}{|\bm h_k^\mathrm{H} \bm w_i|^2}+\sigma_k^2}.
		\end{split}
	\end{equation*}
	\vspace{-10pt}

	The achievable rate of user-$k$ to decode $s_c$ is expressed as
	$R_{c,k}=\log_2(1+\gamma_{c,k})$.
	To ensure that all users can successfully decode $s_c$, the achievable common rate is given by
	$R_c=\min_{k\in\mathcal K}R_{c,k}$.
	Let $r_c$ denote the predefined transmission rate of the common stream, which satisfies $r_c\leq R_c$~\cite{9195771}.
	Following the equal common-rate allocation strategy in~\cite{9195771,11494791}, the common rate allocated to user-$k$ is set as
	$C_k=r_c/K$, such that $\sum_{k\in\mathcal K}C_k=r_c$.
	After SIC, the achievable rate of user-$k$ to decode $s_k$ can be expressed as
	$R_k \!=\!\log_2(1 \!+\!\gamma_k)$.
	Accordingly, the overall achievable rate of user-$k$ is given by $R_{k,{\rm tot}}=C_k+R_k.$
	Then, the overall system sum rate is given by $R_{\text{total}} = r_c +\sum_{k\in\mathcal{K}} R_k$.

	\textit{2) Sensing model:}
	In the proposed ISAC system, the transmitted signal simultaneously serves communication users while sensing a potential target in the region of interest~\cite{li2007mimo}.
	Therefore, we enhance the RIS-reflected sensing beam toward the target direction by jointly designing the BS precoders and RIS phase shifts.
	To achieve this, we propose a beam-gain approximation method, where both common and private streams contribute to the sensing beam. 
	For notational convenience, the common stream is indexed as the $(K \!+\! 1)$-th stream, and the approximation problem is formulated as
	\vspace{-3pt}
	\begin{equation}
		\begin{split}
			\min_{\bm w, \bm \theta, \bm \vartheta}\quad 
			{\sum_{k=1}^{K+1}{|\bm \theta^\mathrm{H}\diag(\bm f_t^\mathrm{H})\mathbf{G}\bm w_k-\delta_k e^{j \vartheta_k}|^2}},
		\end{split}
		\label{eq:beam_gain_approx}
	\end{equation}
	\vspace{-8pt}
	
	\noindent where 
	$\bm f_t$ denotes the steering vector at the RIS with respect to the sensing direction, which is given by 
	$\bm{f}_t = \left[1, e^{-j\pi\sin(\theta^t)\sin(\varphi^t)}, \ldots, e^{-j\pi(N_1-1)\sin(\theta^t)\sin(\varphi^t)}\right]^{\mathrm{T}} \otimes \left[1, e^{-j\pi\cos(\theta^t)}, \ldots, e^{-j\pi(N_2-1)\cos(\theta^t)}\right]^{\mathrm{T}} / \sqrt{N}. $
	$\theta^t$ and $\varphi^t$ denote elevation and azimuth angles of the sensing direction, respectively.
	$\bm{w} \!=\! [\bm{w}_1^\mathrm{T},\ldots,\bm{w}_K^\mathrm{T},\bm{w}_c^\mathrm{T}]^\mathrm{T} \!\in\! \mathbb{C}^{M(K+1)\times 1}.$
	$\delta_k$ denotes the preset desired amplitude gain for the \(k\)-th stream, while $\vartheta_k$ denotes its corresponding beam phase variable.

	\textit{Remark 1:} Since $\mathbb{E}[\mathbf{s}\mathbf{s}^{\text{H}}]\!=\!\mathbf{I}$,
	the cross terms among different data streams vanish after taking the
	expectation, and their power contributions to the average transmit beam
	gain are additive. 
	Thus, equation~\eqref{eq:beam_gain_approx} can be interpreted as a
	stream-wise response-matching criterion for enhancing the average
	transmit beam gain in the target direction, which may improve the
	sensing SNR under fixed receive conditions. Nevertheless, the adopted
	metric does not characterize the conventional beampattern MSE over the
	entire angular domain or complete radar detection performance.

	\vspace{-20pt}
	\subsection{Problem Formulation}
	\vspace{-4pt}
	We aim to jointly optimize the beamforming vector $\bm w$ at the BS, the phase shift vector $\bm \theta$ of RIS, and the beam phase $\bm \vartheta$, with the objective of enhancing the sensing beam gain while satisfying the communication QoS constraints of all users. Accordingly, we formulate the following optimization problem
	\vspace{-15pt}
	\begin{subequations}
		\label{problem_formulation}
		\begin{align}
			\min_{\bm w, \bm \theta, \bm \vartheta}\quad&  \mathcal{F} = {\sum_{k=1}^{K+1}{|\bm \theta^\mathrm{H}\diag(\bm f_t^\mathrm{H})\mathbf{G}\bm w_k-\delta_k e^{j \vartheta_k}|^2}},\label{a}\\
			\text{s.t.} \quad & \mathcal{C}_{\text{BS}}\!:\! \|{\bm{w}}_c\|_2^2 + \sum_{i\in \mathcal{K}}{\|{\bm{w}}_i\|_2^2} \leq P_{\max}, \label{b}\\
			&\mathcal{C}_{\text{QoS}_1}\!:\!\gamma_{c,k} \!=\! \frac{|\bm h_k^\mathrm{H} {\bm{w}}_c|^2}{\sum_{i\in\mathcal{K}}{|\bm h_k^\mathrm{H} {\bm{w}}_i|^2}\!+\!\sigma_k^2} \!\geq\! \gamma^{\text{th}}_c,k \!\in \!\mathcal{K},\label{c}\\
			&\mathcal{C}_{\mathrm{QoS}_2}:
			\gamma_k \!=\!
			\frac{|\bm h_k^{\rm H}\bm w_k|^2}
			{\sum_{\substack{i\in\mathcal K\\ i\ne k}}
				|\bm h_k^{\rm H}\bm w_i|^2 \!+\!\sigma_k^2}
			\!\geq\!\gamma^{\mathrm{th}}_{p,k}, k \!\in\!\mathcal K,\label{d}\\
			& \mathcal{C}_{\text{RIS}}: |\theta_n|=1,n = 1,2,\dots,N,\label{e}
		\end{align}
	\end{subequations}
	\vspace{-15pt}
	
	\noindent where $\bm \vartheta \!=\! [\vartheta_1, \vartheta_2, \dots, \vartheta_K, \vartheta_c]^\mathrm{T} \!\!\in\! \mathbb{R}^{(K+1)\times 1}$.
	${\gamma}_{c}^{\rm th} \!=\!2^{\sum_{i\in\mathcal K}C_i}\!-\!1=2^{r_c}\!-\!1, {\gamma}_{p,k}^{\rm th}\!=\!2^{[R_k^{\min}\!-\! C_k]^+} \!-\!1$.
	A detailed derivation of $\gamma_c^{\mathrm{th}}$ and $\gamma_{p,k}^{\mathrm{th}}$ is provided in~\cite{cheng2026ratecoupled}.
	Constraint \eqref{b} restricts the maximum transmit power at the BS; 
	constraint \eqref{c} ensures the decodability of the common stream at all users, while constraint \eqref{d} guarantees that each user's total achievable rate, composed of the allocated common-rate portion and the private-stream rate, satisfies the minimum rate requirement;
	constraint \eqref{e} imposes the unit modulus RIS constraint.
   	
	Problem \eqref{problem_formulation} is intractable due to the non-convex constraints \eqref{c}, \eqref{d}, and \eqref{e}, as well as the coupling among the optimization variables.
	Existing SCA- or SDR-based methods usually rely on convex optimization tools such as CVX, which incurs high computational complexity.
	\vspace{-11pt}
	\section{Beamforming Optimization}
	\vspace{-6.5pt}
	\label{s3}
	
	In this section, we propose an efficient constraints-separation-based AO (CS-AO) algorithm to iteratively solve problem \eqref{problem_formulation}.
	Specifically, problem \eqref{problem_formulation} is decomposed into three subproblems by alternately optimizing the BS precoder $\bm w$, RIS phase-shift vector $\bm\theta$, and the beam phase vector $\bm\vartheta$.
		
	\vspace{-15pt}
	\subsection{Precoding Vector Optimization}
	\vspace{-5pt}
	With $\bm \theta$ and $\bm \vartheta$ fixed, the problem \eqref{problem_formulation} can be transformed as
	\vspace{-12pt}
	\begin{equation}
		\label{problem_formulation_w}
		\min_{\bm w}\  \mathcal{F}(\bm w)\quad \text{s.t.} \ \mathcal{C}_{\text{BS}},\  \mathcal{C}_{\text{QoS}_1},\  \mathcal{C}_{\text{QoS}_2}.
	\end{equation}
	\vspace{-18pt}
	
	Since the constraints $\mathcal{C}_{\text{QoS}_1}, \mathcal{C}_{\text{QoS}_2}$ are both non-convex, we propose a method to reformulate constraints $\mathcal{C}_{\text{QoS}_1}, \mathcal{C}_{\text{QoS}_2}$ as 
	\vspace{-3pt} 
	\begin{equation}
		\label{problem_constraints}
		\begin{split}
			&2\mathcal{R}\{\hat{\bm w}_c^\mathrm{H}\bm h_k \bm h_k^\mathrm{H}\bm w_c\}-\mathcal{R}\{\hat{\bm w}_c^\mathrm{H}\bm h_k \bm h_k^\mathrm{H}\hat{\bm w}_c\} - \Lambda_{c,k}\geq 0,\\
			&2\mathcal{R}\{\hat{\bm w}_k^\mathrm{H}\bm h_k \bm h_k^\mathrm{H}\bm w_k\}-\mathcal{R}\{\hat{\bm w}_k^\mathrm{H}\bm h_k \bm h_k^\mathrm{H}\hat{\bm w}_k\} - \Lambda_{k}\geq 0,
		\end{split}
	\end{equation}
	\vspace{-10pt}
	
	\noindent where $\Lambda_{c,k} = \gamma^{\text{th}}_c\sigma_k^2 + \gamma^{\text{th}}_c\sum_{i\in\mathcal{K}}|\bm h_k^\mathrm{H}\bm w_i|^2, 
	\Lambda_{k} = \gamma^{\mathrm{th}}_{p,k}\sigma_k^2 + \gamma^{\mathrm{th}}_{p,k}\sum_{i\in\mathcal{K},i\neq k}|\bm h_k^\mathrm{H}\bm w_i|^2. $
	
		
	After the constraint reformulation, problem \eqref{problem_formulation_w} becomes convex. Existing works typically employ CVX to solve such problems. 
	To reduce computational complexity and obtain closed-form updates, we adapt the PDA framework to problem \eqref{problem_formulation_w} through constraint decoupling and approximate it as
	\vspace{-7pt}
	\begin{equation}
		\label{PDA_w}
		\min_{\bm w}
		\mathcal{F}(\bm{w})
		\!+\! \rho \sum_{q=1}^{2}\!\sum_{k=1}^{K}\!
		\operatorname{dist}^2 \!\left(\!\bm w,\mathcal{C}_{\mathrm{QoS}_{q,k}}\!\right)
		\!+\! \rho \operatorname{dist}^2\!\left(\bm w,\mathcal{C}_{\mathrm{BS}}\right)\!,
	\end{equation}
	\vspace{-12pt}
	
	\noindent where $\rho$ is the penalty parameter that controls the weight of the distance-penalty terms,	and $\text{dist}(\mathcal{X},\mathcal{C}_{\mathcal{Y}})$ represents the Euclidean distance from point $\mathcal{X}$ to set $\mathcal{C}_{\mathcal{Y}}$; 
	As $\rho \to \infty$, the optimal solution of problem \eqref{PDA_w} has the same solution as the original problem \eqref{problem_formulation_w}.
	We initialize $\rho$ with a small positive value and gradually increase it during the iterations, which follows homotopy optimization and progressively strengthens the penalty on constraint violations~\cite{10188900}.
	
	Given that $\text{dist}(\mathcal{X},\mathcal{C}_{\mathcal{Y}})$ lacks a closed-form solution and poses computational challenges, majorization-minimization approach is employed to derive a tractable explicit formulation as $\text{dist}(\mathcal{X},\mathcal{C}_{\mathcal{Y}}) \!\leq \!\|\mathcal{X} \!-\! \tilde{\mathcal{X}}_{\mathcal{C}_{\mathcal{Y}}}\|_2$, where $\tilde{\mathcal{X}}_{\mathcal{C}_{\mathcal{Y}}} \!=\! \Pi_{\mathcal{C}_{\mathcal{Y}}}(\bm w)$.
	Consequently, problem \eqref{PDA_w} can be reformulated as the following unconstrained quadratic programming problem.
	\vspace{-6pt}
	\begin{equation}
		\label{PDA_w1}
		\min_{\bm w} \ 
		\mathcal{F}(\bm{w}) \!+\! \rho \bigg(
		\|\bm{w} \!-\! \tilde{\bm{w}}_{\mathrm{BS}}\|_2^2 
		\!+\! \sum_{q=1}^{2}\sum_{k=1}^{K}
		\|\bm{w} \!-\! \tilde{\bm{w}}_{\mathrm{QoS}_{q,k}}\|_2^2
		\bigg),
	\end{equation}
	\vspace{-12pt}
	
	\noindent where $\tilde{\bm w}_{\mathcal{C}_{\mathcal{Y}}} = \Pi_{\mathcal{C}_{\mathcal{Y}}}(\bm{w}) = \arg\min_{\bm y \in \mathcal{C}_{\mathcal{Y}}} \|\bm y - \bm w\|_2^2.$
	
	Through analytical calculations, the projection points required in \eqref{PDA_w1} can be obtained as
	\vspace{-5pt}
	\begin{equation*}
		\label{eq:projection_w_qos1}
		[\tilde{\bm w}_{\text{QoS}_{1,k}}]_j =
		\begin{cases}
			\bm w_j, & \text{if } \gamma_{c,k} \geq \gamma^{\text{th}}_c, \\
			[\bm w_{\text{QoS}_{1,k}}]_j, & \text{otherwise},
		\end{cases}
	\end{equation*}
	\begin{equation*}
		\label{eq:projection_w_qos2}
		[\tilde{\bm w}_{\text{QoS}_{2,k}}]_j =
		\begin{cases}
			\bm w_j, & \text{if } \gamma_k \geq \gamma^{\mathrm{th}}_{p,k}, \\
			[\bm w_{\text{QoS}_{2,k}}]_j, & \text{otherwise},
		\end{cases}
	\end{equation*}
	\begin{equation*}
		\label{eq:projection_w_bs}
		[\tilde{\bm w}_{\text{BS}}]_j =
		\begin{cases}
			\bm w_j, & \text{if } \|\bm w_c\|_2^2+\sum_{i=1}^{K}\|\bm w_i\|_2^2 \leq P_{\max}, \\
			\dfrac{\sqrt{P_{\max}}}{\|\bm w\|_2}\bm w_j, & \text{otherwise},
		\end{cases}
	\end{equation*}
	
	\noindent where
	\vspace{-5pt}
	\begin{equation*}
		[{\bm w}_{\text{QoS}_{1,k}}]_j = 
		\begin{cases}
			\bm w_c + \mu_{c, k} \bm h_k \bm h_k^\mathrm{H} {\hat{\bm w}}_c, & \text{if } j=c, \\
			(\mathbf{I}+\mu_{c, k}\gamma^{\text{th}}_c\bm h_k \bm h_k^\mathrm{H})^{-1}\bm w_k, & \text{if } j=k, \\
			(\mathbf{I}+\mu_{c, k}\gamma^{\text{th}}_c\bm h_k \bm h_k^\mathrm{H})^{-1}\bm w_j, & \text{otherwise},
		\end{cases}
	\end{equation*}
	\vspace{-6pt}
	\begin{equation*}
		[{\bm w}_{\text{QoS}_{2,k}}]_j = 
		\begin{cases}
			\bm w_c, & \text{if } j=c, \\
			\bm w_k + \mu_{p, k} \bm h_k \bm h_k^\mathrm{H} {\hat{\bm w}}_k, & \text{if } j=k, \\
			(\mathbf{I}+\mu_{p, k}\gamma^{\mathrm{th}}_{p,k}\bm h_k \bm h_k^\mathrm{H})^{-1}\bm w_j, & \text{otherwise}.
		\end{cases}
		\vspace{-6pt}
	\end{equation*}
	
	The Lagrange multipliers $\mu_{c, k}$ and $\mu_{p, k}$ must satisfy $2\mathcal{R}\{\hat{\bm w}_c^\mathrm{H} \bm h_k \bm h_k^\mathrm{H} [\bm w_{\text{QoS}_{1,k}}]_c\} - \gamma^{\text{th}}_c\sum_{i=1}^{K}|\bm h_k^\mathrm{H} [\bm w_{\text{QoS}_{1,k}}]_i|^2 - \mathcal{R}\{\hat{\bm w}_c^\mathrm{H}\bm h_k \bm h_k^\mathrm{H} \hat{\bm w}_c\} \!-\! \gamma^{\text{th}}_c\sigma_k^2 \!=\! 0$, $2\mathcal{R}\{\hat{\bm w}_k^\mathrm{H} \bm h_k \bm h_k^\mathrm{H} [\bm w_{\text{QoS}_{2,k}}]_k\} - \gamma^{\mathrm{th}}_{p,k}\sum_{i=1, i \neq k}^{K}|\bm h_k^\mathrm{H} [\bm w_{\text{QoS}_{2,k}}]_i|^2 - \mathcal{R}\{\hat{\bm w}_k^\mathrm{H}\bm h_k \bm h_k^\mathrm{H} \hat{\bm w}_k\} - \gamma^{\mathrm{th}}_{p,k}\sigma_k^2 = 0$, which can be efficiently solved via bisection search.
	The derivation details of $\tilde{\bm w}_{\text{QoS}_{q,k}}$ are provided in Appendix A.
	
	Substituting these projection points into \eqref{PDA_w1}, the final closed-form update of the precoding vector is given by
	\vspace{-5pt}
	\begin{align}
		\bm w_j^\star 
		= &\bigl(\mathbf{G}^\mathrm{H}\bm{\Theta}^\mathrm{H} 
		\bm{f}_t \bm{f}_t^\mathrm{H} 
		\bm{\Theta}\mathbf{G} 
		+ {\rho}(1 + 2K)\mathbf{I}\bigr)^{-1}
		\bigl(\mathbf{G}^\mathrm{H}\bm{\Theta}^\mathrm{H} 
		\bm{f}_t \delta_j e^{j\vartheta_j} 
		\notag\\	
		&+ {\rho}\bigl([\tilde{\bm w}_{\text{BS}}]_j 
		+ \sum_{q=1}^{2}\sum_{k=1}^{K} 
		[\tilde{\bm w}_{\text{QoS}_{q,k}}]_j\bigr)\bigr).
		\label{eq:w_update_closed_form}
	\end{align}
	
	\vspace{-18pt}	
	\subsection{RIS Phase-Shift Vector Optimization}
	With $\bm w$ and $\bm \vartheta$ fixed, the problem \eqref{problem_formulation} can be transformed as
	\vspace{-10pt}
	\begin{equation}
		\label{problem_formulation_theta}
		\min_{\bm \theta}\  \mathcal{F}(\bm \theta)\quad \text{s.t.} \ \mathcal{C}_{\text{QoS}_1},\  \mathcal{C}_{\text{QoS}_2}, \  \mathcal{C}_\text{RIS}.
	\end{equation}
	\vspace{-15pt}	
	
	Following the same method as in \eqref{problem_constraints}, the non-convex QoS constraints 	$\mathcal{C}_{\text{QoS}_1}$ and $\mathcal{C}_{\text{QoS}_2}$ in problem \eqref{problem_formulation_theta} 
	are transformed into $g_1(\bm \theta)\geq 0$ and $g_2(\bm \theta)\geq 0$, respectively, where
	\vspace{-5pt}
	\begin{equation}
		\begin{split}
			g_1(\bm \theta) &= 2\mathcal{R}\{\hat{\bm \theta}^\mathrm{H}\mathbf{Z}_{c,k}{\bm \theta}\} - \mathcal{R}\{\hat{\bm \theta}^\mathrm{H}\mathbf{Z}_{c,k}\hat{\bm \theta}\}-\tilde{\Lambda}_{c,k},\\
			g_2(\bm \theta) &= 2\mathcal{R}\{\hat{\bm \theta}^\mathrm{H}\mathbf{Z}_{k,k}{\bm \theta}\} - \mathcal{R}\{\hat{\bm \theta}^\mathrm{H}\mathbf{Z}_{k,k}\hat{\bm \theta}\}-\tilde{\Lambda}_{k},\\
			\mathbf{Z}_{i,k} &= \diag(\bm f_k^\mathrm{H})\mathbf{G}\bm w_i    \bm w_i^\mathrm{H}\mathbf{G}^\mathrm{H}\diag(\bm f_k),\\
			\tilde{\Lambda}_{c,k} &= \gamma^{\text{th}}_c\sigma_k^2 + \gamma^{\text{th}}_c\sum\nolimits_{i\in\mathcal{K}}\bm \theta^\mathrm{H}\mathbf{Z}_{i,k} \bm \theta, \\
			\tilde{\Lambda}_{k} &= \gamma^{\mathrm{th}}_{p,k}\sigma_k^2 + \gamma^{\mathrm{th}}_{p,k}\sum\nolimits_{i\in\mathcal{K},i\neq k}\bm \theta^\mathrm{H}\mathbf{Z}_{i,k} \bm \theta. 
		\end{split}
	\end{equation}
	
	\vspace{-6pt}
	Following the PDA framework, the problem \eqref{problem_formulation_theta} can be approximated as
	\vspace{-8pt}
	\begin{equation}
		\label{PDA_theta}
		\min_{\bm \theta}
		\mathcal{F}(\bm{\theta})
		+ \rho \bigg(
		\sum_{q=1}^{2}\sum_{k=1}^{K}
		\|\bm{\theta} - \tilde{\bm{\theta}}_{\text{QoS}_{q,k}}\|_2^2
		+ \|\bm{\theta} - \tilde{\bm{\theta}}_{\text{RIS}}\|_2^2
		\bigg),
	\end{equation}

	\vspace{-8pt}
	\noindent where $\tilde{\bm{\theta}}_{\text{QoS}_{q,k}}= \Pi_{\mathcal{C}_{\text{QoS}_{q,k}}}(\bm \theta),q=1,2$, $\tilde{\bm{\theta}}_{\text{RIS}}= \Pi_{\mathcal{C}_{\text{RIS}}}(\bm \theta)$. 

	Subsequently, after some algebraic manipulations, the final closed-form update of RIS phase-shift vector is given by
	\vspace{-6pt}
	\begin{align}
		\bm{\theta}^\star 
		= &\Bigl(\sum_{k=1}^{K+1} 
		\diag(\bm f_t^\mathrm{H}) \mathbf{G}\bm w_k 
		\bm w_k^\mathrm{H} \mathbf{G}^\mathrm{H} 
		\diag(\bm f_t) 
		+ (2K+1)\rho \mathbf{I}\Bigr)^{-1} \notag\\
		&\times \Bigl(\sum_{k=1}^{K+1} 
		\diag(\bm f_t^\mathrm{H}) \mathbf{G}\bm w_k 
		\delta_k e^{-j\vartheta_k} 
		+ \rho \bm c\Bigr),
		\label{eq:theta_update_closed_form}
	\end{align}
	\vspace{-3pt}
	\noindent where 
	$$ \bm c \!=\! \!\sum_{q=1}^{2}\!{\sum_{k=1}^{K}\tilde{\bm{\theta}}_{\text{QoS}_{q,k}}\!\!+\! \tilde{\bm{\theta}}_\text{RIS}},\bm b_{j,k} \!=\! \diag(\bm f_k^\mathrm{H})\mathbf{G} {\bm{w}}_j {\bm{w}}_j^\mathrm{H} \mathbf{G}^\mathrm{H} \diag(\bm f_k),$$ 
	\vspace{-7pt}
	\begin{equation*}
		\tilde{\bm{\theta}}_{\text{QoS}_{1,k}} = \begin{cases}
			\bm{\theta},  \text{if } \frac{|\bm{\theta}^\mathrm{H} \diag(\bm f_k^\mathrm{H})\mathbf{G} {\bm{w}}_c|^2}{\sum_{i=1}^{K}{|\bm{\theta}^\mathrm{H} \diag(\bm f_k^\mathrm{H})\mathbf{G} {\bm{w}}_i|^2}+\sigma_k^2} \geq \gamma^{\text{th}}_c, \\
			\mathbf{A}_{1,k}^{-1}(\bm \theta + \mu_{c,k} \bm b_{c,k}\hat{\bm{\theta}}),  \text{otherwise},
		\end{cases}
	\end{equation*}
	\vspace{-7pt}
	\begin{equation*}
		\tilde{\bm{\theta}}_{\text{QoS}_{2,k}} = \begin{cases}
			\bm{\theta},  \text{if } \frac{|\bm{\theta}^\mathrm{H} \diag(\bm f_k^\mathrm{H})\mathbf{G} {\bm{w}}_k|^2}{\sum_{i=1, i\neq k}^{K}{|\bm{\theta}^\mathrm{H} \diag(\bm f_k^\mathrm{H})\mathbf{G} {\bm{w}}_i|^2}+\sigma_k^2} \geq \gamma^{\mathrm{th}}_{p,k}, \\
			\mathbf{A}_{2,k}^{-1}(\bm \theta + \mu_{p,k} \bm b_{k,k} \hat{\bm{\theta}}),  \text{otherwise},
		\end{cases}
	\end{equation*}
	$$\mathbf{A}_{1,k} = \mathbf{I}+\mu_{c,k}\gamma^{\text{th}}_c\sum_{i=1}^{K} \bm b_{i,k},\mathbf{A}_{2,k} = \mathbf{I}+\mu_{p,k}\gamma^{\mathrm{th}}_{p,k}\sum_{i=1, i\neq k}^{K} \bm b_{i,k},$$
	\begin{equation*}
		[\tilde{\bm{\theta}}_{\text{RIS}}]_n = \frac{\theta_n}{|\theta_n|}, n = 1,2,\dots,N.
	\end{equation*}
	\vspace{-8pt}
	
	It can be observed from \eqref{eq:w_update_closed_form} and \eqref{eq:theta_update_closed_form} that $\rho$ balances the minimization of the original objective and the enforcement of the separated constraints in the closed-form updates.
	
	\vspace{-13pt}
	\subsection{Beam Phase Optimization}
	\vspace{-3pt}
	
	With $\bm w$ and $\bm \theta$ fixed, the problem \eqref{problem_formulation} can be transformed as
	\vspace{-6pt}
	\begin{equation}
		\label{vartheta}
		\begin{split}
			\min_{\vartheta_k} \quad& \mathcal{F}(\bm \vartheta) =\sum_{k=1}^{K+1}{|\bm \theta^\mathrm{H}\diag(\bm f_t^\mathrm{H})\mathbf{G}\bm w_k-\delta_k e^{j \vartheta_k}|^2},\\
			\text{s.t.} \quad & \vartheta_k \in [0,2\pi).
		\end{split}
	\end{equation}
	
	\vspace{-5pt}
	Define $\xi_k \!=\! \bm{\theta}^\mathrm{H} \operatorname{diag}(\bm{f}_t^\mathrm{H}) \mathbf{G} \bm{w}_k$. Since $\left\|\xi_k - \delta_k e^{j \vartheta_k}\right\|_2^2 = |\xi_k|^2 + |\delta_k|^2 - 2\delta_k|\xi_k|\cos(\vartheta_k - \angle\xi_k)$,
	the problem \eqref{vartheta} can be solved by
	\vspace{-3pt}
	\begin{equation}
		\max_{\vartheta_k \in [0,2\pi)} \cos(\vartheta_k - \angle\xi_k).
	\end{equation}
	
	\vspace{-5pt}
	It is evident that the optimal solution is given by $\vartheta_k^\star = \angle \xi_k$.
	The proposed CS-AO algorithm is summarized in Algorithm~1.
	
	\vspace{-15pt}
	\subsection{Convergence and Computational Complexity Analysis}
	\vspace{-3pt}

	According to the PDA convergence result in~\cite{10188900,lange2015pda}, when the inner PDA iterations converge to feasible solutions, the	updates of $\bm w$ and $\bm\theta$ do not increase the
	corresponding block objectives. Moreover, each beam-phase variable
	is updated to its global optimum as $\vartheta_k \!= \! \angle\xi_k$.
	Hence,  $\mathcal{F}(\boldsymbol{w}^{\ell\!+\!1},\boldsymbol{\theta}^{\ell\!+\!1},\boldsymbol{\vartheta}^{\ell\!+\!1}) \!\leq\! \mathcal{F}(\boldsymbol{w}^{\ell\!+\!1},\boldsymbol{\theta}^{\ell\!+\!1},\boldsymbol{\vartheta}^{\ell})
	\!\leq\! \mathcal{F}(\boldsymbol{w}^{\ell\!+\!1},\boldsymbol{\theta}^{\ell},\boldsymbol{\vartheta}^{\ell})\!\leq\!$
	$\mathcal{F}(\boldsymbol{w}^{\ell},\boldsymbol{\theta}^{\ell},\boldsymbol{\vartheta}^{\ell})$.
	Since $\mathcal F\geq0$, the objective-value sequence is monotonically
	non-increasing and therefore convergent.
	The per-iteration complexity of the proposed CS-AO algorithm is 
	$\mathcal{O}\bigl(K^2M^3 \!\mathbin{+}\allowbreak\! K^2NM 
	\!\mathbin{+}\allowbreak\! K^2N^2 \!\mathbin{+}\allowbreak\! KN^3\bigr)$. 
	For comparison, the per-iteration complexities of the SCA and SDR benchmarks are 
	$\mathcal{O}\bigl(K^3M^3 \!\mathbin{+}\allowbreak\! K^2NM 
	\!\mathbin{+}\allowbreak\! K^2N^2 \!\mathbin{+}\allowbreak\! (N+K)^3\bigr)$
	and
	$\mathcal{O}\!\left(\!\!\sqrt{KM}(K^{2}M^{3}\!+\! K^{3}M^{2})\!+\!\sqrt{N \! +\! K}(N^{4}\!+\! KN^{3}\!+\! K^{2}N^{2})\!\right).$
	
	\begin{algorithm}[!t]
		\caption{Proposed CS-AO Algorithm for Problem \eqref{problem_formulation}}
		\label{alg:csao}
		\begin{algorithmic}[1]
			\STATE \textbf{Input:} Initialize $\bm{w}$, $\bm\theta$,
			$\bm\vartheta$, $\rho_{w}>0$, $\rho_{\theta}>0$, $\kappa > 1$; $\ell=1$.
			\REPEAT
			\STATE
			Update $\bm w^{\ell\!+\!1}\!$ via \eqref{eq:w_update_closed_form} with
			$\rho=\rho_{w}$; set
			$\rho_{w}\leftarrow\kappa\rho_{w}$ every $I$ iterations.

			\STATE
			Update $\bm\theta^{\ell+1}$ via \eqref{eq:theta_update_closed_form}	with
			$\rho=\rho_{\theta}$; set
			$\rho_{\theta}\leftarrow\kappa\rho_{\theta}$ every $I$ iterations.
			
			\STATE
			$\vartheta_k^{\ell+1} = \angle \xi_k^{\ell+1} = 
			\angle\left({\bm \theta}^{\ell+1,\mathrm{H}} 
			\operatorname{diag}(\bm{f}_t^\mathrm{H}) 
			\mathbf{G} \bm{w}_k^{\ell+1}\right)$;
			
			\STATE $\ell=\ell+1$;
			
			\UNTIL{$\left|\mathcal{F}^{(\ell)}-\mathcal{F}^{(\ell-1)}\right| < 10^{-4}$ is satisfied.}
		\end{algorithmic}
	\end{algorithm}	
	
	\vspace{-12pt}
	\section{simulation results}
	\vspace{-3pt}
	\label{s4}
	
	In this section, numerical results are presented to demonstrate the performance of the proposed algorithm for RIS-enabled RSMA-ISAC systems.
	The BS and RIS are located at (-20m, 0m, 25m) and (0m, 0m, 0m), respectively. The user-to-RIS distances are randomly distributed in [50, 100] meters.
	For the RIS-to-user channel, we adopt the channel model from \cite{bjornson2020rayleigh1}. The channel vector is modeled as $\bm f_k \sim \mathcal{CN}(0, \beta_k \mathbf{R})$, where $\beta_k = 10^{-3}d_k^{-2}$ and $d_k$ represents the distance between the RIS and user-$k$.
	The large-scale fading of $\mathbf{G}$ is represented by $\mathrm{PL} \!=\! 37.3 \!+ \!22.0\log\left(d\right)$, where $d$ denotes the distance between devices, while the small-scale fading follows the Rician channel model~\cite{wang2021beamforming}. 
	We set $\lambda \!=\! 0.03~\text{m}, \delta_k \!=\!100, \rho_{w} \!=\! \rho_{\theta} \! = \! 1, \kappa \!=\! 2, I \!=\!10,   \gamma^{\text{th}}_c \!=\! \gamma^{\mathrm{th}}_{p,k}\!=\! 1, \sigma_k^2\!=\! -100~\text{dBm}, \theta_t \!=\! 90^\circ, \phi_t \!=\! 45^\circ$. 
	All algorithms are implemented in MATLAB R2024b and run on a desktop with an Intel(R) Core(TM) i5-7500 CPU and 16.0 GB RAM.
		
	\begin{figure}[tbp]
		\centering
		\includegraphics[width=0.23\textwidth]{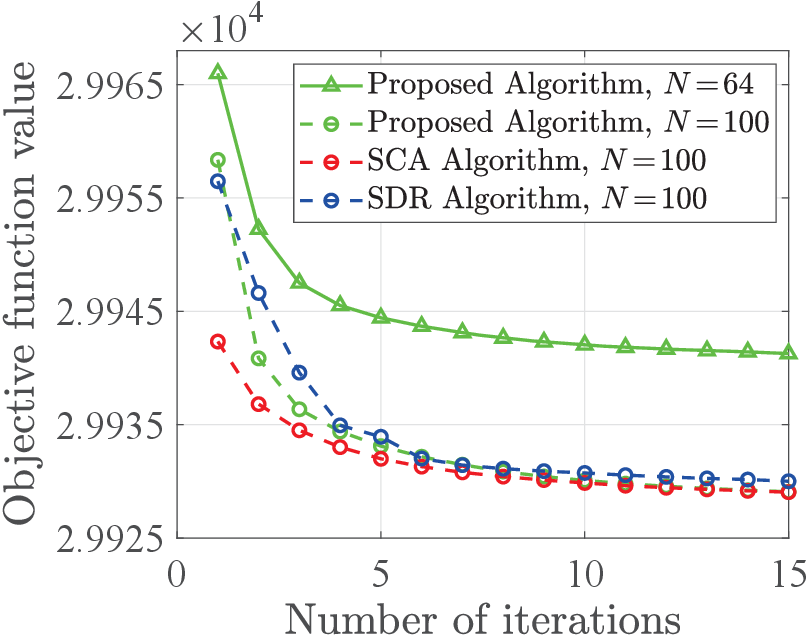}
		\vspace{-5pt}
		\caption{Convergence of different algorithms. $M=10, K=2, P_{\max}= 30~\text{dBm}$.}
		\label{plot_convergence}	
		\vspace{-11pt}
	\end{figure}
	\begin{figure}[tbp]
		\centering
		\subfloat[\small\rmfamily Sensing Beam Gain]{
			\includegraphics[width=0.45\linewidth]{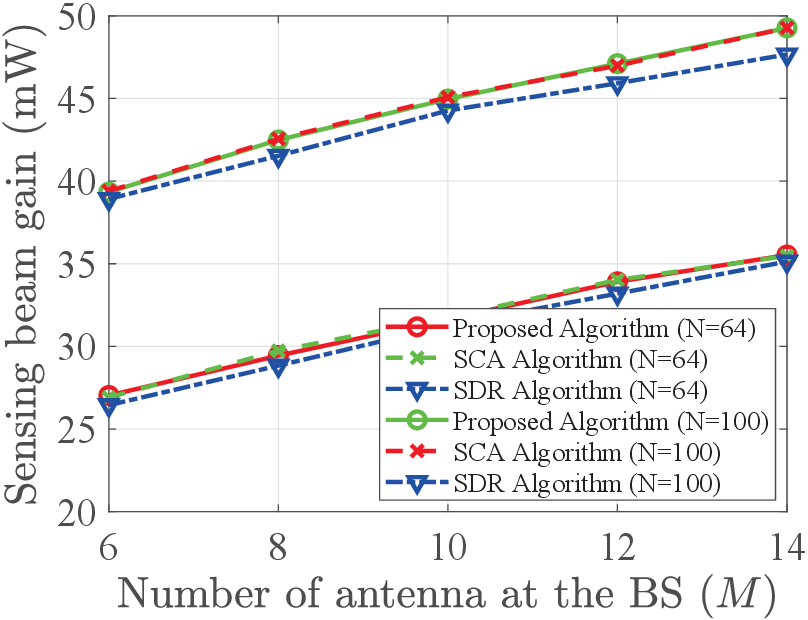}
			\label{comparison_sensing_M}}
		\subfloat[\small\rmfamily Average Runtimes]{
			\includegraphics[width=0.45\linewidth]{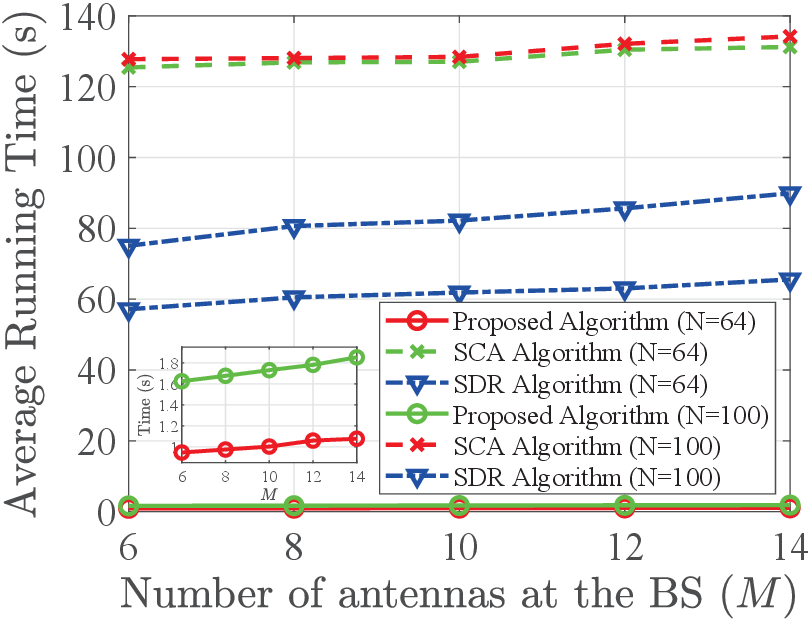}
			\label{comparison_times_M}}
			\vspace{-3pt}
		\caption{Comparison between the proposed algorithm and existing algorithms with different $M$. $K=2,P_{\max}= 30~\text{dBm}$.}
		\label{algorithm_comparison_M}
		\vspace{-11pt}
	\end{figure}
	\begin{figure}[!t]
		\centering
		\subfloat[\small\rmfamily Different BS locations]{
			\includegraphics[width=0.45\linewidth]{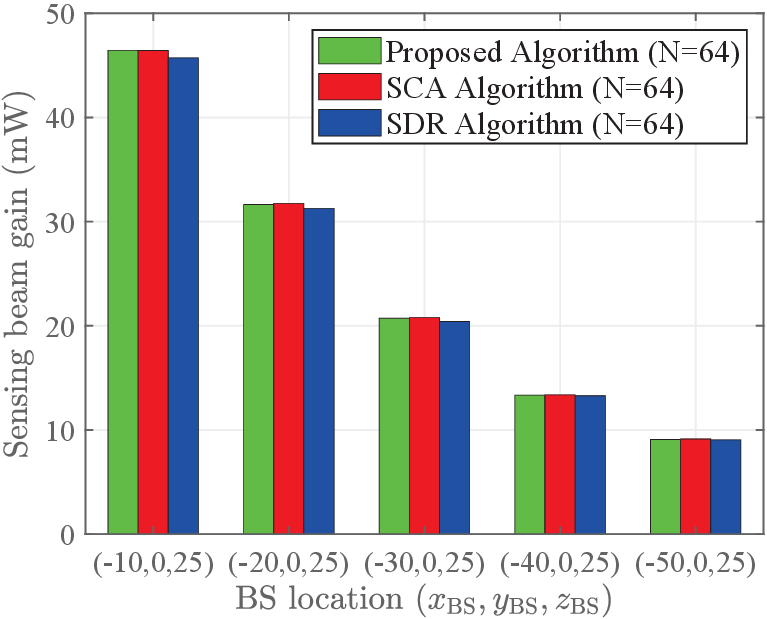}
			\label{comparison_BS}}
		\subfloat[\small\rmfamily Different RIS-user channels]{
			\includegraphics[width=0.45\linewidth]{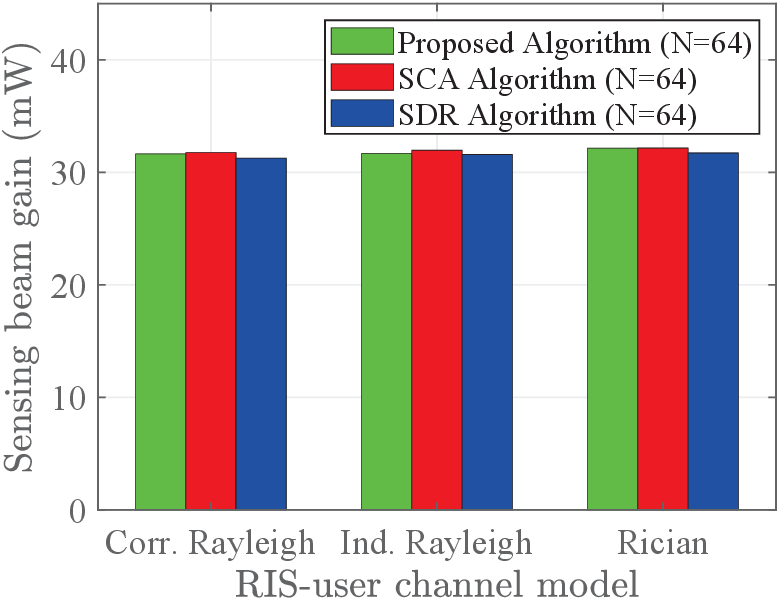}
			\label{comparison_channel}}
		\vspace{-5pt}	
		\caption{Comparison between the proposed algorithm and existing algorithms under different BS locations and RIS-user channel models. \(K=2\), \(M=10\), and \(P_{\max}=30\) dBm.}
		\label{algorithm_comparison_BS_channel}
		\vspace{-3pt}
	\end{figure}

	As shown in Fig.~\ref{plot_convergence}, the proposed CS-AO algorithm converges within a few iterations, which verifies its convergence behavior.
	To verify the effectiveness of our proposed CS-AO algorithm, we compared it with the benchmark SCA and SDR algorithms. 
	Fig.~\ref{algorithm_comparison_M} shows that the proposed algorithm achieves sensing beam gain comparable to the two benchmarks under different numbers of BS antennas $M$, indicating that the closed-form updates can well preserve the sensing beam-gain performance. Meanwhile, the computational efficiency is significantly improved, with more than 120-fold and 50-fold runtime reductions compared with SCA and SDR, respectively. 
	For the runtime results in Fig.~\ref{algorithm_comparison_M}(b), the proposed algorithm achieves standard deviations below $0.2$ s and $0.35$ s for $N \!=\!64$ and $N \!=\!100$, respectively, confirming computational stability.
	
	To verify robustness, we consider different BS locations and RIS-user channel models. As shown in Fig.~\ref{algorithm_comparison_BS_channel}, the sensing beam gain decreases as the BS moves away from the RIS due to increased BS-RIS path loss. 
	Nevertheless, the proposed CS-AO algorithm achieves performance close to the SCA and SDR benchmarks in all considered cases, demonstrating its robustness under different geometries and channel settings.
	To further demonstrate the sensing beam-gain advantage of the proposed RSMA-ISAC scheme, we compare it with the conventional SDMA scheme. 
	
	\begin{figure}[tbp]
		\centering
		\subfloat[{\small\rmfamily Impact of $M$, $K \!= \!8$}]{
			\includegraphics[width=0.43\linewidth]{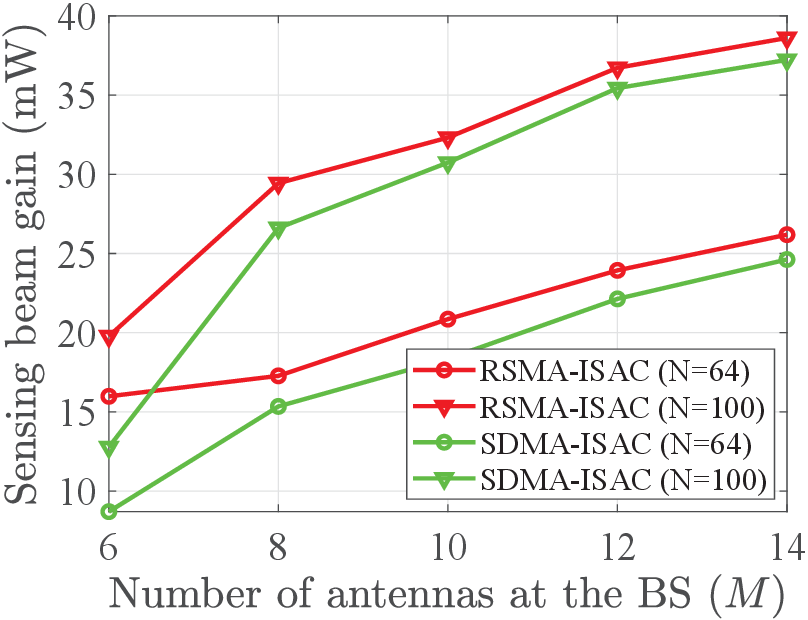}
			\label{RSMA_SDMA_M}}%
		\subfloat[{\small\rmfamily Impact of $K$, $M \!= \!10$}]{
			\includegraphics[width=0.43\linewidth]{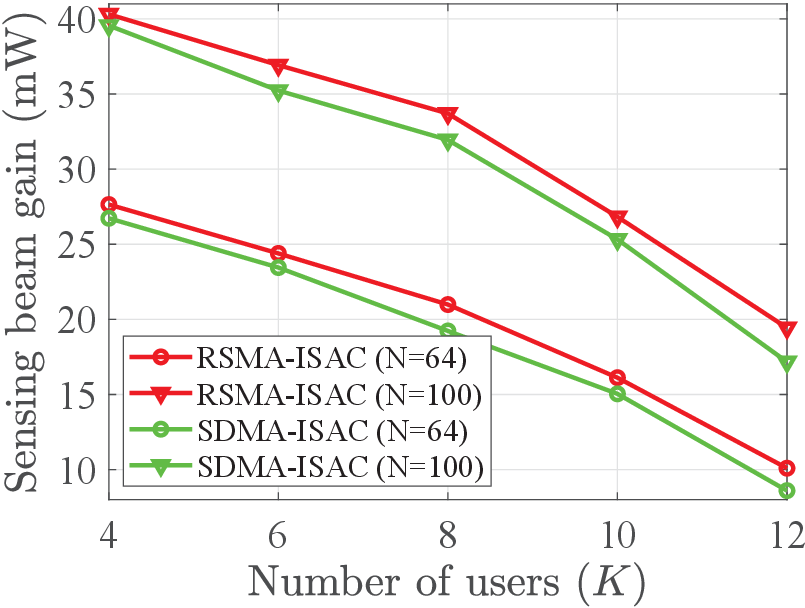}
			\label{RSMA_SDMA_K}}
		\vspace{-3pt}
		\caption{Comparison of sensing beam gain between RSMA and SDMA. $P_{\max}= 30~\text{dBm}$.}
		\label{RSMA_SDMA}
	\end{figure}
	
\begin{figure}[!t]
		\centering
		\includegraphics[width=0.23\textwidth]{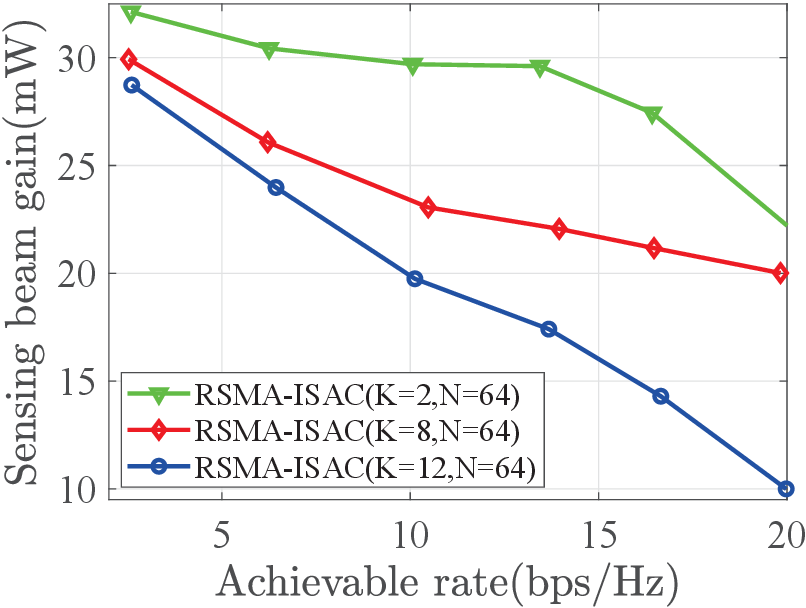}
		\vspace{-5pt}
		\caption{Trade-off performance between achievable rate and sensing beam gain. $M=10, P_{\max}= 30~\text{dBm}$.}
		\label{trade_off}
	\end{figure}

	Fig.~\ref{RSMA_SDMA} shows that the proposed scheme achieves higher system performance, especially in the overloaded regime with $K \! > \! M$.	
	This is because SDMA can no longer sufficiently suppress inter-user interference under overloaded conditions. In contrast, RSMA encodes part of the information into a common stream, thereby providing a more flexible mechanism for inter-user interference management and resulting in better sensing beam gain performance.
	In Fig.~\ref{trade_off}, we simulate the ISAC performance.
	As the number of users or the rate requirement increases, more power is allocated to communication links, leaving less power for sensing and thus reducing the sensing beam gain. 
	This confirms the resource competition between rate enhancement and sensing beam-gain improvement in ISAC systems.
	
	\vspace{-19pt}
	\section{Conclusion}
	\label{s5}
	\vspace{-9pt}
		
	We investigated a sensing-centric design for RIS-enabled RSMA-ISAC systems and proposed a low-complexity structured optimization framework. 
	By introducing a new beam-gain approximation method and a constraint-separation mechanism, the proposed algorithm effectively alleviated the high coupling and non-convexity of the original optimization problem. 
	Simulation results show that the proposed RSMA-ISAC scheme achieves higher sensing beam gain than SDMA, while the CS-AO algorithm maintains comparable sensing beam-gain performance to SCA and SDR with significantly lower runtime. 
	Future research may consider more general and practical models and constraints, such as imperfect channel state information and cooperative multi-RIS deployment, to broaden its applicability to intelligent wireless sensing and communication networks.

	\vspace{-13pt}
	\section*{Appendix A: Proof of $\tilde{\bm w}_{\text{QoS}_{1,k}}$}
	\label{appendices_0}
	\vspace{-5pt}
	The projection $\tilde{\bm w}_{\text{QoS}_{1,k}}$ can be obtained by solving the following optimization problem
	\vspace{-8pt}
	\begin{equation}
		\label{AAA}
		\begin{split}
			\min_{\bm w} \  \|\bm w - \bm y\|_2^2, \quad
			\text{s.t.} \quad  g_3{(\bm w)} \geq 0,
			\vspace{-8pt}
		\end{split}
	\end{equation}
	
	\vspace{-5pt}
	\noindent where $\bm w = [\bm w_1^\mathrm{T},\dots,\bm w_K^\mathrm{T},\bm w_{c}^\mathrm{T}]^\mathrm{T}$ denotes the projection point.
	$g_3{(\bm w)} = 2\mathcal{R}\{\hat{\bm w}_c^\mathrm{H}\bm h_k \bm h_k^\mathrm{H}\bm w_c\}-\mathcal{R}\{\hat{\bm w}_c^\mathrm{H}\bm h_k \bm h_k^\mathrm{H}\hat{\bm w}_c\} - \Lambda_{c,k}\geq 0$
	
	Then, we derive the KKT conditions for problem \eqref{AAA} as
	\vspace{-5pt}
	\begin{equation*}
		\begin{split}
			&\partial L/\partial \bm w_c^* 
				\! = \! \bm w_c \! -\!  \bm y_c \! -\!\mu_{c,k} \bm h_k \bm h_k^\mathrm{H} \hat{\bm w}_c \! = \!0,      \mu_{c,k} \!\geq\! 0, g_3(\bm w) \!\geq\! 0,\\
			&\partial L/\partial \bm w_k^* 
				\! = \! \bm w_k \! -\! \bm y_k \!+\!\mu_{c,k} \gamma^{\text{th}}_c \bm h_k \bm h_k^\mathrm{H} \bm w_k \! = \!0, \mu_{c,k} g_3(\bm w) \!=\! 0,\\
			&\partial L/\partial \bm w_{j}^*  
				\! = \! \bm w_j \! -\! \bm y_j \!+\! \mu_{c,k} \gamma^{\text{th}}_c \bm h_k \bm h_k^\mathrm{H} \bm w_j \! = \!0, j \!\in\! \mathcal{K}, j \!\neq\! k,\\
		\end{split}
	\end{equation*}
	
	\vspace{-5pt}
	\noindent where $\mu_{c, k}$ is a dual variable. 
	From stationarity conditions and Woodbury matrix identity, we obtain the closed-form solution
	\vspace{-9pt}
	\begin{equation}
		[{\bm w}]_j = 
		\begin{cases}
			\bm y_c + \mu_{c, k} \bm h_k \bm h_k^\mathrm{H} {\hat{\bm w}}_c, & \text{if } j=c, \\
			(\mathbf{I}+\mu_{c, k}\gamma^{\text{th}}_c\bm h_k \bm h_k^\mathrm{H})^{-1}\bm y_k, & \text{if } j=k, \\
			(\mathbf{I}+\mu_{c, k}\gamma^{\text{th}}_c\bm h_k \bm h_k^\mathrm{H})^{-1}\bm y_j, & \text{otherwise}.
		\end{cases}
	\end{equation}
	
	\vspace{-6pt}	
	By complementary slackness, $\mu_{c,k}\!=\!0$ if the constraint is inactive.
	For $\mu_{c,k}\!>\!0$, the constraint is active, and $\mu_{c,k}$ satisfies
	$\gamma_c^{\mathrm{th}}\sum_{i=1}^{K}\left|\bm h_k^{\mathrm H}[\bm w]_i\right|^2 \!+\! \gamma_c^{\mathrm{th}}\sigma_k^2 \!+\!\mathcal{R}\!\left\{\widehat{\bm w}_c^{\mathrm H}\bm h_k\bm h_k^{\mathrm H}\widehat{\bm w}_c\right\}\!-\!2\mathcal{R}\!\left\{\widehat{\bm w}_c^{\mathrm H}\bm h_k\bm h_k^{\mathrm H}[\bm w]_c\right\}\!=\!0$. 
	The projection $\tilde{\bm w}_{\mathrm{QoS}_{2,k}}$ follows analogous steps, with the detailed derivation provided in~\cite{cheng2026ratecoupled}.
	
	\bibliographystyle{IEEEtran}
	\vspace{-15pt}
	\bibliography{myref}
	\vspace{-15pt}
\end{document}

%% file: sym_marco.tex
\newcommand{\diag}{\mathrm{diag}}